\documentstyle[aps,prd,preprint,tighten,amssymb,eqsecnum,newlfont]{revtex}
\newcommand{\beq}{\begin{equation}}
\newcommand{\eeq}{\end{equation}}
\newcommand{\bea}{\begin{eqnarray}}
\newcommand{\eea}{\end{eqnarray}}
\newcommand{\eps}{{\varepsilon}}
\begin{document}
\preprint{
\begin{tabular}{r}
UWThPh-2000-05\\
February 2000
\end{tabular}
}
\draft
\title{Is it possible to determine the S--factor of the hep process
from a laboratory experiment?}
\author{
W.M. Alberico$^{\mathrm{a}}$,
J. Bernab\'eu$^{\mathrm{b}}$,
S.M. Bilenky$^{\mathrm{c,a}}$ and
W. Grimus$^{\mathrm{d}}$
}
\address{$^{\mathrm{a}}$INFN, Sezione di Torino,\\
and Dipartimento di Fisica Teorica, Universit\`a di Torino,\\
Via P. Giuria 1, 10125 Torino, Italy}
\address{$^{\mathrm{b}}$Departament de F\'{\i}sica Te\`orica,
Universitat de Val\`encia,\\
E--46100 Burjassot, Val\`encia, Spain}
\address{$^{\mathrm{c}}$Joint Institute for Nuclear Research, Dubna, Russia}
\address{$^{\mathrm{d}}$Institute for Theoretical Physics, University
of Vienna,\\
Boltzmanngasse 5, A--1090 Vienna, Austria}
\maketitle
\begin{abstract}
We discuss the problem of solar hep neutrinos originating from the reaction
$p + {^3\mathrm{He}} \to {^4\mathrm{He}} + e^+ + \nu_e$ 
and obtain a relation between the
astrophysical S--factor of the hep process and the cross section of the process
$e^- + {^4\mbox{He}} \to {^3\mbox{H}} + n + \nu_e$ near threshold. 
The relation is based on the isotopic invariance of strong interactions. 
 The measurement of the latter cross section
would allow to obtain  experimental information on $S(hep)$,
the value of which, at the moment, is known only from theoretical
calculations.
\end{abstract}


\renewcommand{\thefootnote}{\fnsymbol{footnote}}

\section{Introduction}

The reaction
\beq\label{main}
p + {^3\mathrm{He}} \to {^4\mathrm{He}} +e^+ + \nu_e \,,
\eeq
the so--called hep reaction, is the source of solar neutrinos with the
highest energies (up to 18.8~MeV). It is expected that the cross section
of this process at solar energies ($\lesssim 30$~keV) is extremely small and
cannot be measured in laboratory conditions \cite{BK-hep}. 
The most detailed calculations of the cross section of the 
hep process were done in Refs.~\cite{Carl,Schiav}.

The process (\ref{main}) proceeds via a Gamov--Teller transition 
(a change in isospin from 1 to 0 is involved). In ref.\cite{Carl,Schiav}
there are two reasons for the strong suppression of the hep cross section.
The first one is due to the fact that 
the matrix element of the process vanishes in the
allowed approximation if only the main $s$--state components of $^4$He
and $^3$He wave functions are taken into account \cite{werntz}.
The second reason lies in strong cancellations between different matrix
elements of the weak nuclear current: it was found in Ref.~\cite{Schiav} 
that the contribution to the 
matrix element of the hep process from the diagrams with $\pi$ and $\rho$ 
meson exchange currents (in which the transition of
a nucleon into the $\Delta$ isobar is also taken into account) is comparable
in modulus with the contribution of the one--body current, but has opposite
sign. As a result there is a cancellation of the contributions of these 
two terms and the calculated cross section is about 5 times smaller than 
the cross section predicted by the one--nucleon term only.

The result of the calculation also depends on the two-body nuclear potential:
by using two different, typical NN--potentials, for the S--factor of hep 
process in Ref.~\cite{Schiav} the following values were obtained:
\bea
S_1(hep) & = & 1.44 \times 10^{-20} \mathrm{~keV~b} \,,
\nonumber\\
S_2(hep) & = & 3.14 \times 10^{-20} \mathrm{~keV~b} \,.
\nonumber
\eea
The average between these two values, 
\beq
S_0(hep)= 2.3 \times 10^{-20} \;\mathrm{keV~b} \,,
\label{Sfactor}
\eeq
is used in the Standard Solar Model (SSM) \cite{BP-95}. If the value of 
the astrophysical S--factor of the hep process is given by Eq.(\ref{Sfactor})
then the total flux of hep--neutrinos \cite{BP-98},
\beq
\Phi(hep) = 2.1 \times 10^{3}\; \mathrm{cm}^{-2}\,\mathrm{s}^{-1},
\label{Phihep}
\eeq
is more than three orders of magnitude smaller than the flux of $^8$B
neutrinos,
\[ \Phi({^8{\mathrm B}}) = 5.15 \times
\left( 1.00^{+0.19}_{-0.14} \right) \times 10^{6} \;
\mathrm{cm}^{-2}\,\mathrm{s}^{-1},\]
and hep neutrinos give a negligible contribution to the event rates 
observed in solar neutrino experiments. Let us stress, however, that
the value (\ref{Sfactor}) of $S_0(hep)$ is the result of a very
complicated and model-dependent calculations.

The recent interest in hep neutrinos was triggered by the results of the
Super-Kamiokande experiment\cite{sol,nakahata,suzuki} in which
the spectrum of recoil electrons in the solar-neutrino-induced process
\[ \nu + e^- \to \nu + e^-  \]
was measured.
The spectrum of neutrinos from the decay 
${^8\mbox{B}} \to {^8\mbox{Be}} + e^+ + \nu_e$ is 
determined by weak interactions and well known. The recoil electron 
spectrum measured in the Super-Kamiokande experiment is in agreement 
with the predicted spectrum in the whole energy range starting 
from 5.5~MeV, with the exception of the highest energy region, in
which two data points with large errors 
(from the bins 13.5 -- 14~MeV and 14 -- 20~MeV)
are above the prediction \cite{sol,nakahata}.

The Super-Kamiokande recoil electron spectrum 
can be fitted with the assumption that there is 
no distortion of the spectrum ($\chi^2=24.3$ at 17 d.o.f.)
\cite{nakahata,suzuki}. In Ref.~\cite{BK-hep} attention was payed, 
however, to the fact that the high energy points in the Super-Kamiokande data
could be due to the contribution of hep neutrinos \cite{escribano}. 
If one considers $S(hep)$ as a free parameter, then from the fit of the data 
(504 days of Super-Kamiokande) in Ref.~\cite{BK-hep} in the hypothesis of no 
oscillations $S(hep)/S_0(hep)=26$ was obtained. In a more recent 
fit \cite{nakahata} (825 days of Super-Kamiokande) it was found 
$S(hep)/S_0(hep)=16$ ($\chi^2=19.5$ at 16 d.o.f.).

Let us notice that the distortion of the recoil electron spectrum 
could also be due
to the MSW effect or to vacuum oscillations (VO). The largest enhancement
of the high energy part of the spectrum is expected for the VO solution.
By fitting the data with $S(hep)=S_0(hep)$ in the case of VO solution it was 
found $\sin^2 2\theta=0.79$, $\Delta m^2=4.3 \times 10^{-10}$~eV$^2$ 
($\chi^2=44.1$ at 35 d.o.f.) \cite{nakahata} (see also 
Refs.~\cite{BK-hep,GG,BKS-SNO}).

The investigation of the problem of hep neutrinos in the solar neutrino 
Super-Kamiokande experiment will be continued. The results of a
new measurement of the spectrum of solar $\nu$'s in the region 
$E \ge 5$~MeV
will be soon available from the SNO experiment \cite{SNO}. 
In this experiment the spectrum of solar $\nu_e$'s will be determined from 
the measurement of the electron spectrum in the process 
$\nu_e + d \to e^- + p + p$.

One of the central theoretical problems connected with hep neutrinos is the 
astrophysical S--factor of the hep process \cite{fiorentini}. 
``The most important unsolved 
problem in theoretical nuclear physics related to solar neutrinos is the 
range of values allowed by fundamental physics for the hep production 
cross section''(Bahcall \cite{Bahcall}).
In this letter we consider the possibility to determine $S(hep)$ from 
experimental data. We will obtain here a relation between
$S(hep)$ and the total cross section of the process
\beq
e^- + {^4\mathrm{He}} \to {^3\mathrm{H}} + n + \nu_e 
\label{inverse}
\eeq
near threshold. The relation we obtain is based on the isotopic invariance 
of the strong interactions (we neglect the Coulomb interaction in the region
of nuclear forces). From the existing nuclear data it follows that the
violation of isotopic invariance for light nuclei cannot be larger than 
$10-20\%$.\footnote{This estimate follows from nuclear mass differences,
mirror nuclei spectra and so on.}

From the point of view of a possible investigation of the process 
(\ref{inverse}) at small energies, we would like to stress two points:
\begin{enumerate}
\item
The cross section of the process (\ref{inverse}) does not contain the
Coulomb penetration factor, which suppresses at small energies
the cross sections of processes with initial particles having charges 
of equal sign.
\item
There exist electron accelerators (microtrons) which allow to obtain high 
intensity electron beams in the range of energies which are appropriate 
for the investigation of the process (\ref{inverse}).
\end{enumerate}

\section{The relation between $S(\lowercase{hep})$ and 
$\sigma(\lowercase{e}^- \: {^4\mathrm{H\lowercase{e}}} \to 
{^3\mathrm{H}}\, \lowercase{n}\, \nu_{\lowercase{e}})$}

Let us start by considering the process (\ref{main})
at small solar energies ($\lesssim 30$~keV).
In Ref.~\cite{Fowler} the general arguments are given that at small
energies the cross sections of the reactions with charged initial
particles have the form
\beq
\sigma(E) =\frac{1}{E} e^{-2\pi\eta} S(E) \,.
\label{sigma}
\eeq
Here $E$ is the kinetic energy of the initial particles in the C.M.
system and
\beq
\eta=\frac{Z_1 Z_2 e^2}{v} \,,
\label{eta}
\eeq
where $Z_1e$, $Z_2e$ are the charges of the initial particles,
$v=\sqrt{2E/\mu}$ is their relative velocity and $\mu$ is the reduced
mass. In the expression (\ref{sigma}),
\beq
P\equiv e^{-2\pi\eta}
\label{coul}
\eeq
is the probability of penetration of the incident particle through
the Coulomb barrier \cite{gamow} and the factor $S(E)$ is determined
mostly by strong interactions.
If there are no resonances at small energies, the function $S(E)$
depends very weakly on the energy $E$.

The relation (\ref{sigma}) with $S\simeq \mathrm{const.}$ allows to describe
the existing low energy data and is used for the extrapolation of
laboratory data to the energy region which is relevant for solar reactions
(see, for example, Ref.~\cite{BahcAdel}). 
Our further considerations will be based on this relation.

The standard weak interaction Hamiltonian density is given by
\beq
{\cal H}_I=\frac{G_F}{\sqrt{2}}{\bar\nu}_e\gamma^\alpha
(1-\gamma_5)e j_\alpha + \mathrm{h.c.} \,,
\label{hamilt}
\eeq
where the hadronic $V-A$ current 
$j_\alpha= j_\alpha^1 -i j_\alpha^2\equiv j_\alpha^{1-i2}$ is the
``minus'' component of the isovector $j^a_\alpha$ ($a=1,2,3$).

For the matrix element of the process (\ref{main}) we have
\bea
\langle f|S|i \rangle & = & -i\frac{G_F}{\sqrt{2}}\frac{1}{(2\pi)^3}
\frac{1}{\sqrt{4k_0k_0'}}\ell^{\alpha}(k,k')
\nonumber\\
& \times &
\int d^4x\, e^{iq\cdot x} \langle {\mathrm{^4He}}|T\left( j_\alpha(x)
e^{-i\int d^4y {\cal H}^0_I(y)} \right)|p\, {\mathrm{^3He}} \rangle \, .
\label{smatrix}
\eea
Here $\ell^{\alpha}(k,k')= \bar{u}(k')\gamma^{\alpha}(1-\gamma_5)v(k)$ is
the matrix element of the weak leptonic current,
$q=k+k'$ ($k$ and $k'$ being the momenta of the $e^+$ and $\nu_e$,
respectively) and ${\cal H}^0_I={\cal H}^h_I+{\cal H}^{em}_I$ is the
Hamiltonian density of strong (${\cal H}^h_I$) and electromagnetic
(${\cal H}^{em}_I$) interactions.

Let us first consider only that part of the matrix element of the process
(\ref{main}) which is determined by the strong interactions and gives
the major contribution to the $S$--factor.
Neglecting the Coulomb interaction in the region of nuclear forces, we have
for the hadronic part of the matrix element (\ref{smatrix}) 
\beq
\langle f|S|i \rangle = 
-i\frac{G_F}{\sqrt{2}}\frac{1}{(2\pi)^3}\frac{1}{\sqrt{4k_0k_0'}}
\ell^\alpha(k,k')
\langle {\mathrm{^4He}}|J^{(-)}_\alpha(0)|p\, {^3\mathrm{He}} \rangle
(2\pi)^4\delta(P'-P) \,,
\label{strong2}
\eeq
where $J^{(-)}_\alpha(x)\equiv J_\alpha^{1-i2}(x)$ is the hadronic weak $V-A$
current in the {\em Heisenberg representation} and $P$ ($P'$)
is the total four-momentum of the initial (final) states.
It is evident  that $\langle {\mathrm{^4He}}|J^{(-)}_\alpha(0)|p\,
{^3\mathrm{He}} \rangle $ includes {\it all} possible contributions
coming from strong interactions (for example, in addition to the one--body
nucleonic current, also two--body exchange currents, effects of the
$\Delta$ isobar and so on).

Using the charge symmetry of strong interactions we have
\bea
&&\langle {^4\mathrm{He}}|J^{1-i2}_\alpha|p\, {^3\mathrm{He}} \rangle =
\langle {^4\mathrm{He}}|{\cal U}^{-1}{\cal U}J^{1-i2}_\alpha
{\cal U}^{-1}{\cal U}|p\, {^3\mathrm{He}} \rangle
\nonumber\\
&&=- \langle {^4\mathrm{He}}|J^{1+i2}_\alpha|n\, {^3\mathrm{H}} \rangle =
-\langle n\, {^3\mathrm{H}}|J^{1-i2}_\alpha| {^4\mathrm{He}} \rangle^*.
\label{charsym}
\eea
Here ${\cal U}=\exp\{i\pi T_2\}$ is the unitary operator of
 rotation by an angle $\pi$ around the second axis in isospace.
In Eq.(\ref{charsym}) we took into account that
\bea
{\cal U}J_\alpha^{1-i2}{\cal U}^{-1} & = & - J_\alpha^{1+i2} \,,
\nonumber\\
{\cal U}\, |p\, {^3\mathrm{He}} \rangle & = & |n\, {^3\mathrm{H}} \rangle \,,
\label{varia}\\
{\cal U}\, |{^4\mathrm{He}} \rangle & = & |{^4\mathrm{He}} \rangle \,.
\nonumber
\eea

Thus the hadronic part of the matrix elements of the process
$e^- + {^4{\mathrm{He}}} \to {^3\mathrm{H}} + n + \nu_e$
is connected with the matrix element of the hadronic weak current for
 the hep process by the simple charge symmetry relation (\ref{charsym}).
With the help of Eq.(\ref{charsym}) we will obtain a relation which
connects the $S$--factor of the hep process (\ref{main}) with the
cross section of the process (\ref{inverse}).

Let us continue considering the hep process.
In the region of small energies we are interested in, only the contribution
of the $s$--wave of the initial $p-^3$He system is relevant. Taking into 
account the Coulomb interaction between the initial $p$ and $^3$He,
for the total cross section of the hep process in the center of mass 
system we have
\bea
\sigma(hep) && = 
 \frac{(2\pi)^4}{v}\, \frac{G_F^2}{2}\,
\frac{1}{4}\sum_\mathrm{spins}
\int \frac{d^3k}{2k_0} \int \frac{d^3k'}{2k_0'} 
\label{totsig} \\
&&\times\left|\ell^\alpha 
\langle {^4\mathrm{He}}|J^{(-)}_\alpha |p\, {^3\mathrm{He}} \rangle
\right|^2 \delta(k_0+k_0'-\Delta)\,\frac{|\psi^{(+)}_{\vec p}(0)|^2}
{|\psi_{\vec p}|^2} \,.
\nonumber
\eea
Here $\Delta = m_p + m_{^3\mathrm{He}} - m_{^4\mathrm{He}} = 19.284$~MeV
(see, e.g., Ref.~\cite{firestone} for the values of the nuclear masses),
$v=\sqrt{2E/\mu}$ is the relative velocity, $E$ being the initial energy 
and $\mu$ the reduced mass of the  $p-{^3}$He system,
$\psi^{(+)}_{\vec p}(0)$ is the Coulomb wave function of the initial
$p-{^3}$He system at $r=0$ and
$\psi_{\vec p}({\vec r}\,)=\exp (i{\vec p}\cdot{\vec r})/(2\pi)^{3/2}$.
Notice that in (\ref{totsig}) we have neglected the small recoil energy
of ${^4\mathrm{He}}$  and that the factor $1/4$
is due to averaging over the spin states of the initial particles.

For the hep process the non-relativistic expression
\beq
\frac{|\psi^{(+)}_{\vec p}(0)|^2}{|\psi_{\vec p}|^2}=
\frac{2\pi\eta}{e^{2\pi\eta}-1}
\label{coulcorr}
\eeq
holds, where
\beq
\eta=\frac{2e^2}{v}\simeq 8.66\frac{1}{\sqrt{E[{\mathrm{keV}}]}}\, .
\label{eta1}
\eeq
In the region of small energies $2\pi\eta\gg 1$, we have, for the Coulomb
factor (\ref{coulcorr}), 
\beq
\frac{|\psi^{(+)}_{\vec p}(0)|^2}{|\psi_{\vec p}|^2}\simeq
2\pi\eta\, e^{-2\pi\eta} \,.
\label{etapprox}
\eeq
 For the small energies we are interested in, we have $| {\vec q}\, | R\ll 1$, 
$R$ being the radius of nuclear forces. Moreover, the parity of the 
initial and final nuclear states is the same and in the matrix element
the linear term in the expansion of $\exp{ \{i{\vec q}\cdot{\vec x}\} }$ 
vanishes.  Thus, in Eq.(\ref{smatrix}) we can put 
$e^{-i{\vec q}\cdot{\vec x}} \simeq 1$ (allowed transition). In this
approximation, the matrix element of the hadronic current does not
depend on ${\vec q}$.\footnote{
This independence of the matrix element upon ${\vec q}$ is confirmed by the
detailed  calculations made in Refs.~\cite{Carl,Schiav}, in which not only 
the one--nucleon term but also two--body terms
due to the exchange of $\pi$ and $\rho$ mesons were taken into account.
Nevertheless we think that further investigation of the $q$--dependence
of the nuclear matrix element is an important and interesting issue.}
Moreover, in that region the matrix element 
$\langle {^4\mathrm{He}}|J_\alpha |p\, {^3\mathrm{H}} \rangle$  does not
depend on the relative momentum ${\vec p}$ of the initial particles
either. Since only the axial vector current contributes to the matrix
element (Gamow--Teller transition) we obtain
\bea
\lefteqn{\frac{1}{4k_0k_0'} \sum_\mathrm{spins} \left| \ell^\alpha
\langle {^4\mathrm{He}}|J^{(-)}_\alpha |p\, {^3\mathrm{He}} 
\rangle\right|^2 =} 
\nonumber \\ &&
\frac{1}{4k_0k_0'} \sum_\mathrm{spins} \ell^i{\ell^k}^{*}\delta_{ik}
\frac{1}{3}\left|
\langle {^4\mathrm{He}}|{\vec J}\,^{(-)}|p\, {^3\mathrm{He}} 
\rangle\right|^2 = 
2 \left(1-\frac{1}{3}\frac{{\vec k}\cdot{\vec k'}}{k_0 k_0'}\right) \,
\sum_\mathrm{spins} 
\left| \langle{^4\mathrm{He}}|{\vec J}\,^{(-)}|p\,{^3\mathrm{He}}
\rangle\right|^2 \,.
\label{detail}
\eea
It is obvious that the second term in the last equality does not give
a contribution to the total cross section. Taking into account the Coulomb
interaction of the final $e^+$ and $^4$He, from Eqs.(\ref{totsig}),
(\ref{eta}) and (\ref{detail}) we obtain the total cross section of the hep
process 
\beq
\sigma(hep)=
\frac{(2\pi)^7}{E}\, G_F^2 e^2 m_{e}^5\mu
\sum_\mathrm{spins} \left|
\langle {^4\mathrm{He}}|{\vec J}\,^{(-)}|p\,{^3\mathrm{He}} \rangle\right|^2
f(\eps_0)\, e^{-4\pi e^2/v} \,. 
\label{total}
\eeq
Here $f(\eps_0)$ is given by
\beq
f(\eps_0) = \int_{1}^{\eps_0} F(-2,\eps) (\eps_0-\eps)^2
\sqrt{\eps^2-1}\, \eps\, d\eps \,,
\label{ff}
\eeq
where $\eps=k_0/m_e$, $\eps_0=\Delta/m_e$ and $F(-2,\eps)$ is the Fermi
function (ratio of the modulus squared of the positron wave function in
the Coulomb field of the final nucleus, calculated at $r=R$, to the
modulus squared of the plane wave). Tables of the Fermi function are
given in Ref.~\cite{feshbach}. For the hep reaction we have
$\eps_0 \simeq 37.7$. Using the approximation $F(-2,\eps) \simeq 1$ valid
for small $Z$ and large positron energies, we obtain \cite{feshbach}
$f(\eps_0) \simeq \eps_0^5/30 \simeq 2.55 \times 10^6$.

Finally, from Eqs.(\ref{sigma}) and (\ref{total}), we derive the
following expression for the S--factor for the hep process:
\beq
S(hep) = (2\pi)^7 G_F^2\, e^2 \mu\, m_e^5 f(\eps_0)
\sum_\mathrm{spins} \left| 
\langle {^4\mathrm{He}}|{\vec J}\,^{(-)} |p\,{^3\mathrm{He}} 
\rangle\right|^2 \,.
\label{Sfac}
\eeq

Let us consider now the process (\ref{inverse}). Neglecting the
electromagnetic interaction of hadrons, for the matrix
element of the process we get an expression similar to Eq.(\ref{strong2}):
\beq
\langle f|S|i \rangle = -i\frac{G_F}{\sqrt{2}}\,
\frac{1}{(2\pi)^3}\frac{1}{\sqrt{4k_0k_0'}}\ell^\alpha(k,k')
\langle n\, {\mathrm{^3H}}|J^{(-)}_\alpha(0)|{^4\mathrm{He}} \rangle
(2\pi)^4\delta(P'-P) \,,
\label{strong3}
\eeq
where 
$\ell^{\alpha}(k,k')={\bar u}(k')\gamma^{\alpha}(1-\gamma_5) u(k)$ and
$k$ and $k'$ are the four--momenta of $e^-$ and $\nu_e$, respectively.

The threshold for the process (\ref{inverse}) is given by (see, e.g., 
Ref.~\cite{firestone} for a table of nuclear masses)
\beq
E_\mathrm{th} = \frac{(m_n+m_{^3\mathrm{H}})^2 - m^2_{^4\mathrm{He}}}%
{2 m_{^4\mathrm{He}}} \simeq 21.167 \; \mbox{MeV}.
\label{thres}
\eeq
We will consider the process (\ref{inverse}) at electron energies close
to the threshold energy $E_\mathrm{th}$. For the total cross section we
have
\bea
&&{\sigma(e^-\: {^4\mathrm{He}} \to {^3\mathrm{H}}\, n\, \nu_e) =}
\label{totsig1}\\
&&
\frac{1}{v_e}\,(2\pi)^4\,\frac{G_F^2}{2}\, 
F( 2, E_e/m_e )\,
\frac{1}{2E_e}\int \frac{d^3 k'}{2k_0'} \int d^3 p \int d^3 p' \frac{1}{2}
\sum_\mathrm{spins} \left| 
\langle n\, {^3\mathrm{H}}|J^{(-)}_{\alpha}|{^4\mathrm{He}} 
\rangle \ell^{\alpha}\right|^2 \delta(P'-P) \,,
\nonumber
\eea
where $E_e \equiv k_0$ is the electron energy, $v_e$ the velocity of the
electron, $p'$ and $p$ are the total and relative momenta, respectively, 
of the $n-{^3\mathrm{H}}$ system 
and the Fermi function $F(2,E_e/m_e)$ takes into account the Coulomb
interaction between the initial $e^-$ and ${^4\mathrm{He}}$.
Within the same approximations we have used in the derivation of
the relation (\ref{Sfac}), the cross section for the process (\ref{inverse})
turns out to be
\bea
\lefteqn{\sigma(e^-\: {^4\mathrm{He}} \to {^3\mathrm{H}}\, n\, \nu_e ) =}
\label{totsig2} \\ &&\qquad 
\frac{32}{105}\, (2\pi)^6\, G_F^2 \sum_\mathrm{spins} \left| 
\langle n\, {^3\mathrm{H}}|{\vec J}\,^{(-)}|{^4\mathrm{He}} \rangle \right|^2
(E_e-E_\mathrm{th})^{7/2}\, \mu
\sqrt{2\mu}\, F( 2, E_e/m_e ) \,,
\nonumber
\eea
where $\mu$ is the reduced mass of the $n-{^3\mathrm{H}}$ system (704.1
MeV) which we identify numerically with that of the
$p-{^3\mathrm{He}}$ system (703.3 MeV),
in agreement with our assumption of 
isotopic invariance of the strong interactions.
Using now the isotopic relation (\ref{charsym}), which connects the hadronic 
parts of the S--matrix elements of the processes (\ref{main}) and 
(\ref{inverse}), 
we obtain the following relation between the total cross section of the
process (\ref{inverse}) and the astrophysical S--factor of
the hep--process:
\beq
\sigma(e^-\: {^4\mathrm{He}} \to {^3\mathrm{H}}\, n\, \nu_e) = 
\frac{32}{105}\frac{1}{(2\pi)e^2} \sqrt{\frac{2\mu}{m_e}} \,
\frac{F( 2, E_e/m_e )}{f(\eps_0)}
\left(\frac{E_e-E_\mathrm{th}}{m_e}\right)^{7/2} \frac{S(hep)}{m_e} \,.
\label{final}
\eeq
For the case of electron energies $E_e > 20$~MeV, we 
can set the Fermi function equal to one. 
Then we get for the cross section of the reaction (\ref{inverse})
\beq
\sigma(e^-\: {^4\mathrm{He}} \to {^3\mathrm{H}}\, n\, \nu_e) \simeq
0.62 \times 10^{-50} \; \mathrm{cm}^2 \times 
\left(\frac{E_e-E_\mathrm{th}}{m_e}\right)^{7/2}
\frac{S(hep)}{S_0(hep)} \,.
\label{num}
\eeq

The relation (\ref{final}) allows to determine the astrophysical S--factor
of the hep process 
$p + {^3\mathrm{He}} \to {^4\mathrm{He}} +e^+ + \nu_e$
directly from the measurements of the cross section of the process
$e^- + {^4\mathrm{He}} \to {^3\mathrm{H}} + n + \nu_e$
near threshold.\footnote{Note 
that the kinetic energy of the $n-{^3\mathrm{H}}$
system in its C.M. system, which corresponds to the kinetic C.M. 
system energy of $p - {^3\mathrm{He}}$ in the hep reaction, is given by 
$E(n\,{^3\mathrm{H}}) \simeq E_e - E_\mathrm{th} - k'_0$\,,
where $k_0'$ is the final neutrino energy.}
If, for example, $S(hep)=20 S_0(hep)$, a representative value among the
ones indicated by the analysis of the Super--Kamiokande 
data\cite{BK-hep,suzuki,GG,BKS-SNO}, then at 
$E_e-E_{\mathrm{th}}\simeq 10$~MeV, 
for the cross section of the process (\ref{inverse}) we get the
value
\[\sigma(e^-\: {^4\mathrm{He}} \to {^3\mathrm{H}}\, n\, \nu_e) 
\simeq 4.0 \times 10^{-45}$~cm$^2.\] 
Obviously, the cross section of the process (\ref{inverse})
is small (weak interactions and small energies). However, taking into account 
the importance  of obtaining direct experimental information  on $S(hep)$,
it is worthwhile from our point of view to consider the possibility of
performing such a measurement.
One can use the advantage of high intensity beams of low energy electrons
from microtrons and the possibility to detect the process (\ref{inverse}) 
by radiochemical or mass-spectrometric methods.

\section{Conclusions}

The problem of hep neutrinos is one of the most important issues in solar
neutrino physics and in solar neutrino oscillations. The 
direct measurement of the cross section of the process (\ref{main})
at solar energies does not seem to be possible with the 
present techniques and the calculation of the cross section of the hep 
process is a very complicated problem.

In this paper we have obtained a relation between the astrophysical 
S--factor of the hep process and the total cross section of the process
(\ref{inverse}) near the threshold 
($E_\mathrm{th}\simeq 21.167$~MeV). The relation 
is based on the isotopic invariance
of strong interactions. The measurement of the cross section of this process
near threshold would allow 
to determine the major hadronic part of $S(hep)$. Though the smallness of the
cross section (\ref{num}) precludes going so close to the threshold such that
$E_e - E_\mathrm{th}$ is of the order of the temperature in the solar core,
measuring the process (\ref{inverse}) at the more realistic energies
$E_e - E_\mathrm{th} \sim 10$ MeV might allow to extrapolate 
$S(hep)$ to smaller energies. 
Therefore, we believe that it is worthwhile to consider
the possibility of measuring the cross section (\ref{num})
at microtron facilities.

\acknowledgments

 We are indebted to F. von Feilitzsch, D. Harrach, A. Molinari,
T. Ohlsson and M. Rho for very useful discussions and suggestions on
the problems considered in this work.\\

 The research has been supported
in part by CICYT, under the Grant AEN/99-0692, by Italian MURST and 
INFN research funds.


\begin{thebibliography}{99}

\bibitem{BK-hep}
J.N. Bahcall and P.I. Krastev, %hep-ph/9807525
Phys. Lett. B \textbf{436}, 243 (1998).

\bibitem{Carl} 
J. Carlson \textit{et al.}, 
Phys. Rev. C \textbf{44}, 619 (1991).

\bibitem{Schiav} 
R. Schiavilla \textit{et al.},
Phys. Rev. C \textbf{45}, 2628 (1992).

\bibitem{werntz}
C. Werntz and J.G. Brennan, 
Phys. Rev. \textbf{157}, 759 (1967).

\bibitem{BP-95}
J.N. Bahcall and M.N. Pinsonneault,
Rev. Mod. Phys. \textbf{67}, 871 (1995).

\bibitem{BP-98}
J.N. Bahcall, S. Basu and M.N. Pinsonneault,
Phys. Lett. B \textbf{433}, 1 (1998).

\bibitem{sol}
Super-Kamiokande Coll., M.B. Smy,
Talk presented at DPF'99, hep-ex/9903034.

\bibitem{nakahata}
Super-Kamiokande Coll., M. Nakahata,
Talk presented at TAUP99, www page\\
http://taup99.in2p3.fr/TAUP99/.

\bibitem{suzuki}
Y. Suzuki,
Talk presented at \textit{Lepton-Photon '99}, www page\\
http://www-sldnt.slac.stanford.edu/lp99/db/program.asp.

\bibitem{escribano}
R. Escribano \textit{et al.}, Phys. Lett. B \textbf{444}, 397 (1998).

\bibitem{GG}
M.C. Gonzalez-Garcia \textit{et al.}, hep-ph/9906469.

\bibitem{BKS-SNO}
J.N. Bahcall, P.I. Krastev and A.Yu. Smirnov, hep-ph/9911248.

\bibitem{SNO}
SNO Coll., \textit{The Sudbury Neutrino Observatory}, nucl-ex/9910016.

\bibitem{fiorentini}
G. Fiorentini \textit{et al.},
Phys. Lett B \textbf{444}, 387 (1998).

\bibitem{Bahcall} 
J.N. Bahcall,
in: \textit{Neutrino Mixing}, ed. W.M. Alberico (World Scientific,
Singapore, 2000).

\bibitem{Fowler} 
W.A. Fowler, Lectures in Theor. Phys., vol. VI, p. 379 (Colorado Press,
Boulder, 1964).

\bibitem{gamow}
G. Gamow, Phys. Rev. \textbf{53}, 595 (1938).

\bibitem{BahcAdel} 
E. Adelberger \textit{et al.},
Rev. Mod. Phys. \textbf{70}, 1265 (1998).

\bibitem{firestone}
R.B. Firestone, \textit{Table of Isotopes} 
(John Wiley \& Sons., Inc., New York, 1996).

\bibitem{feshbach}
A. DeShalit and H. Feshbach,
\textit{Theoretical Nuclear Physics} 
(John Wiley \& Son, Inc., New York, 1974). 

\end{thebibliography}
\end{document}